\documentclass[12pt]{iopart}
\usepackage{graphicx}
\usepackage{xcolor}
 \usepackage{hhline}
 \usepackage{iopams}
 \usepackage{citesort}
 \expandafter\let\csname equation*\endcsname\relax
\expandafter\let\csname endequation*\endcsname\relax
\usepackage{amsmath}

\begin{document}
\title[Predicting transitions in cooperation levels from network connectivity]{Predicting transitions in cooperation levels from network connectivity}
\author{A. Zhuk$^{1,+}$,  I. Sendi\~na-Nadal$^{2,3,+}$, I. Leyva$^{2,3,+}$, D. Musatov$^{1,4,5}$, A.M. Raigorodskii$^{1,5,6,7}$,  M. Perc$^{8,9,10,11}$, and S. Boccaletti$^{1,2,12}$}
\address{$^1$Moscow Institute of Physics and Technology (National   Research University), 9 Institutskiy per., Dolgoprudny, Moscow   Region, 141701, Russia }
\address{$^2$Universidad Rey Juan Carlos, Calle Tulip\'an s/n, 28933 M\'ostoles, Madrid, Spain}
\address{$^3$Center for Biomedical Technology, Universidad Polit\'ecnica de Madrid, 28223 Pozuelo de Alarc\'on, Madrid, Spain}
\address{$^4$Russian Academy of National Economy and Public Administration, pr. Vernadskogo, 84, Moscow, 119606, Russia }
\address{$^5$ Caucasus Mathematical Center at Adyghe State University,  ul. Pervomaiskaya, 208, Maykop, 385000, Russia}
\address{$^{6}$Mechanics and Mathematics Faculty, Moscow State University, Leninskie Gory, 1, Moscow, 119991, Russia}
\address{$^{7}$Institute of Mathematics and Computer Science, Buryat State University, ul. Ranzhurova, 5, Ulan-Ude, 670000, Russia}
\address{$^8$Faculty of Natural Sciences and Mathematics, University of Maribor, Koro{\v s}ka cesta 160, 2000 Maribor, Slovenia}
\address{$^9$Department of Medical Research, China Medical University Hospital, China Medical University, 404332 Taichung, Taiwan}
\address{$^{10}$Alma Mater Europaea, Slovenska ulica 17, 2000 Maribor, Slovenia}
\address{$^{11}$Complexity Science Hub Vienna, Josefst{\"a}dterstra{\ss}e 39, 1080 Vienna, Austria}
\address{$^{12}$CNR - Institute of Complex Systems, Via Madonna del Piano
  10, I-50019 Sesto Fiorentino, Italy}

\eads{\mailto{irene.sendina@urjc.es},\mailto{inmaculada.leyva@urjc.es}\\
$^+$These Authors contributed equally } 

\begin{abstract}
Networks determine our social circles and the way we cooperate with others. We know that topological features like hubs and degree assortativity affect
 cooperation, and we know that cooperation is favoured if the benefit of the altruistic act divided by the cost exceeds the average number of neighbours.
 However, a simple rule that would predict cooperation transitions on an arbitrary network has not yet been presented. Here we show that the unique sequence
 of degrees in a network can be used to predict at which game parameters major shifts in the level of cooperation can be expected, including phase transitions from absorbing to mixed strategy phases. We use the evolutionary prisoner's dilemma game on random and scale-free
 networks to demonstrate the prediction, as well as its limitations and possible pitfalls. We observe good agreements between the predictions and the results obtained with concurrent and Monte Carlo methods for the update of the strategies, thus providing a simple and fast way to estimate the outcome of evolutionary social dilemmas on arbitrary networks without the need of actually playing the game.
\end{abstract}

\submitto{\NJP}

\section{Introduction}

In 1992 Nowak and May observed that cooperators form compact clusters and can thereby withstand invading defectors in an iterated prisoner's dilemma game on a square lattice~\cite{nowak_n92b}. This was a fascinating discovery because cooperators should have died out in agreement with the Nash equilibrium of the game~\cite{nash_pnas50,nash_am51}. This would, in fact, always happen in a well-mixed population, but not on a network~\cite{szabo_pr07,perc_bs10,wang_z_epjb15}. The fact that the structure of a network can positively affect the evolution of cooperation is today known as {\it network reciprocity}~\cite{nowak_n92b,santos_prl05}, and it constitutes one of five key mechanisms for cooperators to survive social dilemmas~\cite{nowak_s06}.

With the advent of network science at the turn of the 21st century~\cite{albert_rmp02,newman_siamr03,boccaletti_pr06}, the field of evolutionary games on networks rapidly gained on popularity, and various complex networks have been studied for their impact on the evolution of cooperation, including scale-free~\cite{santos_pnas06,gomez-gardenes_prl07,rong_pre07,poncela_njp07,masuda_prsb07,tomassini_ijmpc07,szolnoki_pa08,assenza_pre08,santos_n08,poncela_pre11,tanimoto_pre12,amaral2018heterogeneous2}, small-world~\cite{abramson_pre01,kim_bj_pre02,masuda_pla03,santos_pre05,tomassini_pre06,fu_epjb07,vukov_pre08}, hierarchical \cite{vukov_pre05,lee_s_prl11}, coevolving~\cite{zimmermann_pre04,pacheco_prl06,fu_pa07,szolnoki_epl08,fu_pre09,szolnoki_njp09}, and empirical social networks~\cite{holme_pre03,guimera03,lozano_ploso08}. Later on, interdependent and multilayer networks have emerged as an important new paradigm in network science~\cite{boccaletti_pr14,kivela_jcn14}, and these were also considered prominently in cooperation research~\cite{gomez-gardenes_pre12,jiang_ll_srep13,wang_pre14a,battiston2017determinants,shen_epl18,fotouhi2018conjoining,shi_nody19,duh_19,sun2020aspiration}, including the discovery of interdependent network reciprocity as an extension of the traditional case~\cite{wang_z_srep13}. Most recently, the focus has been shifting once more, this time to higher-order networks~\cite{battiston2020networks}, where evolutionary dynamics has been considered~\cite{burgio2020evolution,alvarez-rodriguez_nhb21,boccaletti21}.

However, despite the wealth of research concerning the evolution of cooperation on networks, fundamental results have been relatively scarce. { Tanimoto \cite{Tanimoto2007} and Wang et al.~\cite{Wang2015} introduced  universal scaling parameters to account for the strength of a dilemma disrupting the promotion of cooperation  that can evaluate the outcome of a game independently of the reciprocity mechanisms, and showed how these mechanisms relax the dilemma strength \cite{Hiromu2018}}. Ohtsuki et al.~\cite{ohtsuki_n06}, for instance, presented a simple rule for the evolution of cooperation on graphs and social networks, proving that natural selection favours cooperation if the benefit of the altruistic act, divided by the cost, exceeds the average number of neighbours. Moreover, Allen et al.~\cite{Allen_2017} derived a general formula for pairwise games under weak selection that applies to any graph or social network.

In this paper, we derive a simple conjecture for cooperation transitions in evolutionary social dilemmas based on the unique sequence of degrees in a network. As we will show, the conjecture predicts well at which game parameters important shifts in the level of cooperation can be expected, including phase transitions from absorbing to mixed strategy phases~\cite{javarone2016conformity,javarone_epjb16}.

In what follows, we first present the derivation of the conjecture and then proceed with showing the results obtained for the evolutionary prisoner's dilemma game on random and scale-free networks. We conclude by discussing the implications of our research to optimise large-scale simulations of evolutionary processes on complex networks and the possibilities for the generalisation of the conjecture to related subjects.

\section{Game and conjecture formulation}\label{framework}
We start by considering a population of $N$ agents
playing the Prisoner's Dilemma (PD) game
\cite{rapoport1989prisoner}. At each step of the game, agents are allowed to play different strategies, that is, a node can cooperate with some neighbours and defect with others \cite{diverse2020}, { also known as the mixed strategy system with different offers \cite{Kokubo2015,Tanimoto2017}.}

Connections between players follow the network structure prescribed by the adjacency matrix $A=(a_{uv})$, where $a_{uv}=1$ if players $u$ and $v$ are linked, and $a_{uv}=0$ otherwise.

The payoffs in the PD game are extracted from the following matrix:

\begin{equation}
    { PD}(b, c) =
\begin{array}{c|cc}
 &{\rm C} & {\rm D}\\ \hhline{-|--}
{\rm C} &b-c  & -c\\ \hhline{~|~}
{\rm D} & b & 0 \\
\end{array}
\end{equation}
which means that a cooperator (C) altruistically loses an amount $c>0$ when playing with a defector (D), who wins an amount $b>c$ in such interaction. An interplay between two cooperators results in a mutual benefit of $b-c$, whereas playing between two defectors results in a null interchange. Note that a defector never looses.

Without loss of generality, one can take $c=1$ {($b>1$)} and re-scale the payoff matrix as:
\begin{equation}
    P(b) = { PD}(b, 1) / b =
    \begin{pmatrix}
    1 - 1/b & -1/b \\
    1 & 0 \\
    \end{pmatrix}.
    \label{Pb}
\end{equation}
{ According to the universal scaling for the dilemma strength introduced in Ref.~\cite{Wang2015}, $D'_g=D'_r=1/(b-1)$, the parameter $b$ measures the dilemma weakness.}

For interacting players $u$ and $v$, one can define the strategy variable $x_{uv}$ as

\begin{equation}
    x_{uv} =  \left\{
    \begin{array}{lll}
        1 & \mbox{if} & u \; \mbox{cooperates with } v \\
        0 & \mbox{if} & u \; \mbox{defects } v
    \end{array}\right. \nonumber
\end{equation}
and the strategy vector as $\overline{x}_{uv}=\big( x_{uv}, 1-x_{uv}\big)$. In each round of the game, the  payoff $g_{uv}$ earned by each player $u$ from its interaction with $v$ is:
\begin{equation}
   g_{uv} = \overline{x}_{uv}P(b)\overline{x}_{vu}^T. \label{guv}
 \end{equation}
 The average payoff $g_u$ collected by player $u$ after an interaction round with its $k_u$ neighbors is thus given by
\begin{equation}
    g_{u} = \frac{1}{k_u}\sum_{v \in N(u)}g_{uv},\label{gu}
\end{equation}
being $N(u)$ the set of neighbors of $u$.

Evolution of the game is designed by means of the following update rule: at each time step, all pairs of linked players engage in the PD game and their payoffs $g_u$ are updated according to Eq.~(\ref{gu}). Each player $u$ then compares its performance with its neighborhood. Let us refer to the  neighbor of $u$ with the highest payoff as $w$, and as  $x_{wu}$ its strategy adopted against $u$. Then, $u$ updates its strategy $x_{uv}$ imitating its best neighbor strategy $x_{wu}$ with probability
\begin{equation}
    p(\alpha,\Delta g_u) =  \frac1{1+e^{-\frac{\Delta g_{u}}{b\alpha}}},\label{prob}
\end{equation}
which is a Fermi function \cite{fermi}, i.e. a sigmoid function of the free parameter $\alpha$ (playing the role of a temperature) and the payoff difference $\Delta g_{u} = g_w-g_u$ (playing the role of an energy). Therefore, the update rule of the strategy $x_{uv}(t)$ of $u$ at time $t$ is given by
\begin{equation}
    x_{uv}(t+1)= \left\{
    \begin{array}{ll}
			x_{uv}(t) & \mbox{with probability } 1 - p\\
            x_{wu} (t)  & \mbox{with probability  } p,
      \end{array} \right.
      \label{eq:link_strategy}
\end{equation}
that is, player $u$ imitates, at time $t+1$, the strategy that its best neighbor { $w$} played against it in the previous step with $k_u p$ of its neighbors, and keeps the same strategy { is actually playing } with { each one of }the rest.

One can define the node outcoming  cooperation rate as
\begin{equation}
    \rho_u = \frac1{k_u} \sum_{v \in N(u)} x_{uv},\label{rhou}
\end{equation}
and the incoming cooperation rate as
\begin{equation}
    \tau_u = \frac1{k_u}{\sum_{v \in N(u)} x_{vu}},\label{tauu}
\end{equation}
and the macroscopic cooperation frequency of the whole network as
\begin{equation}
    \rho = \frac1{N}{\sum_{u=1}^N \rho_u}.
\end{equation}

Note that defection is very profitable when the benefit $b$ is close to $1$, while when $b\to \infty$ the profitability of defection reduces. As a consequence, in well-mixed populations, defection is the dominant trait driving cooperation to extinction. However, it is known that, in structured populations, cooperation can survive provided the appropriate conditions \cite{Ohtsuki2006,santos_prl05}.

In this context, we now explore how the transition from defection to cooperation depends on the connectivity, as the dilemma weakness $b$ increases. To do this, let us consider two linked players $u$ and $v$ with strategy vectors $\overline{x}_{uv}=(x_{uv},1-x_{uv})$ and $\overline{x}_{vu}=(x_{vu},1-x_{vu})$. Using Eq.~(\ref{Pb}), one can write the payoff of the first player from engaging the second as
\begin{equation}
    \overline{x}_{uv}P(b)\overline{x}_{vu}^T = x_{vu} - \frac{x_{uv}}{b}.
\end{equation}

Therefore, the total payoff accumulated by $u$ is

\begin{equation}
    g_u = \frac1{k_u}{\sum_{v \in N(u)} \overline{x}_{uv} P(b) \overline{x}_{vu}^T} = \frac1{k_u}{\sum_{v \in N(u)} \left(x_{vu} - \frac{x_{uv}}{b}\right)} ,
\end{equation}
or, equivalently, using definitions (\ref{rhou}) and (\ref{tauu})

\begin{equation}
g_u = \tau_u - \frac{\rho_u}{b}.\label{gusimp}
\end{equation}

Let us focus now on the probability of changing strategy, which becomes 50\% when the payoff difference $\Delta g_u=0$ in Eq.~(\ref{prob}). Whenever at some point during the iterated game a player $u$ reaches  such a null payoff difference with respect to its best performing neighbor $w$, the transition to keep the actual strategies $x_{uv}$ or to change to $x_{wu}$ is critical in the sense that it occurs for a particular value of the parameter $b$. This implies that, evaluating payoffs of $u$ and $w$ using Eq.~(\ref{gusimp}),
\begin{equation}
   \Delta g_u=\tau_w-\tau_u -\frac{1}{b*}(\rho_w-\rho_u)=0,
\end{equation}
a critical value of $b^*$ is evaluated as

\begin{equation}
    b^* = \displaystyle\frac{\rho_w - \rho_u}{\tau_w - \tau_u} =
    \frac{
    \frac 1 {k_w} \sum_{z\in N(w)} x_{wz} - \frac 1 {k_u} \sum_{v\in N(u)} x_{uv}
    }{ \frac 1 {k_w} \sum_{z \in N(w)} x_{zw}
    -\frac 1 {k_u} \sum_{v \in N(u)} x_{vu}.
    }
\end{equation}
From this relationship, one can formulate a simple conjecture based on the connectivity pattern to locate the values of $b$ at which cooperation is likely to transit to higher (or lower) frequencies: given a PD game on a graph $G$, all phase transitions are located at points of the form

\begin{equation}
    b^*=\frac{p/k_v - q/k_u}{r / k_v - s / k_u} = \frac{p k_u - q k_v}{r k_u - s k_v},\label{linkconjec}
\end{equation}
where $0 \le p,r\le k_v$, $0\le q,s\le k_u$, are integers (such that the denominator is non-zero).

Note that the set of $b^*$ values resulting from Eq.~(\ref{linkconjec}) tells us about the possible locations of a phase transition. Equation ~(\ref{linkconjec}) entirely depends on the unique sequence of degrees present in the graph and, therefore, it alone allows predicting the transitions without actually playing the game.
However, the multiplicity of each $b^*$
does depend on the particular degree distribution, and one can extract important information about the transitions from this degeneracy, as shown in the next section.

While in the game framework described so far, a player can  perform different strategies with its neighbors, in the more classic scenario, players adopt a single role in each run. We will refers to this updating rule as {\it node strategy}, in opposition to the {\it link strategy} defined by Eq. (\ref{eq:link_strategy}).  It is straightforward to obtain the equivalent equation to that of Eq.~(\ref{linkconjec}) in this latter case:

 \begin{equation}
     b^* =\frac{1}{r/ k_v - s/k_u} = \frac{k_u k_v}{r k_v - s k_u}, \label{nodeconjec}
 \end{equation}
where $0\le r \le k_u$ and $ 0\le s \le k_v$ are integers (such that denominator is non-zero) and
Eq.~(\ref{nodeconjec}) indicates that the possible locations for a phase transition in a node strategy situation is reduced with respect to the link strategy choice.

Notice that the two previous conjectures for link ({ mixed}) and node strategies, Eqs.~(\ref{linkconjec}) and (\ref{nodeconjec}) respectively, depend on the particular form of the payoff matrix. Therefore, the relationship between game parameters and topology has to be recomputed if we change the dilemma, as the critical points $\Delta g_u=0$ depend on them.  For example, in order to illustrate this with a case extensively investigated in the literature, we chose the dilemma's matrix with no cost 
\begin{equation}
    P'(b)=
    \begin{pmatrix}
    1  & 0 \\
    b & 0 \\
    \end{pmatrix}.
    \label{Pbp}
\end{equation}
 as in Refs.\cite{santos_prl05,nowak_n92a,perc_njp09}. { Note that parameter $b$ in Eq.~(\ref{Pbp}) now plays the role of a temptation to defect. In this case, the corresponding universal dilemma strengths \cite{Wang2015} are $D'_g=b$ and $D'_r=0$, therefore, parameter $b$ directly accounts for the strength of the dilemma.}
 
 Using a similar procedure as the one leading to Eq.~(\ref{nodeconjec})  one can obtain an equivalent conjecture to locate the phase transitions for the dilemma $P'$ above,
\begin{equation}
b^*=\frac{n}{m}
\label{eq:bstarPnc}
\end{equation}
where $n$ and $m$ are integers such that $1\le m\le n\le \max(k)$, and $\max(k)$ is the maximum degree present in the network. 


\section{Results}
\begin{figure}
    \centering
    \includegraphics[width=0.5\textwidth]{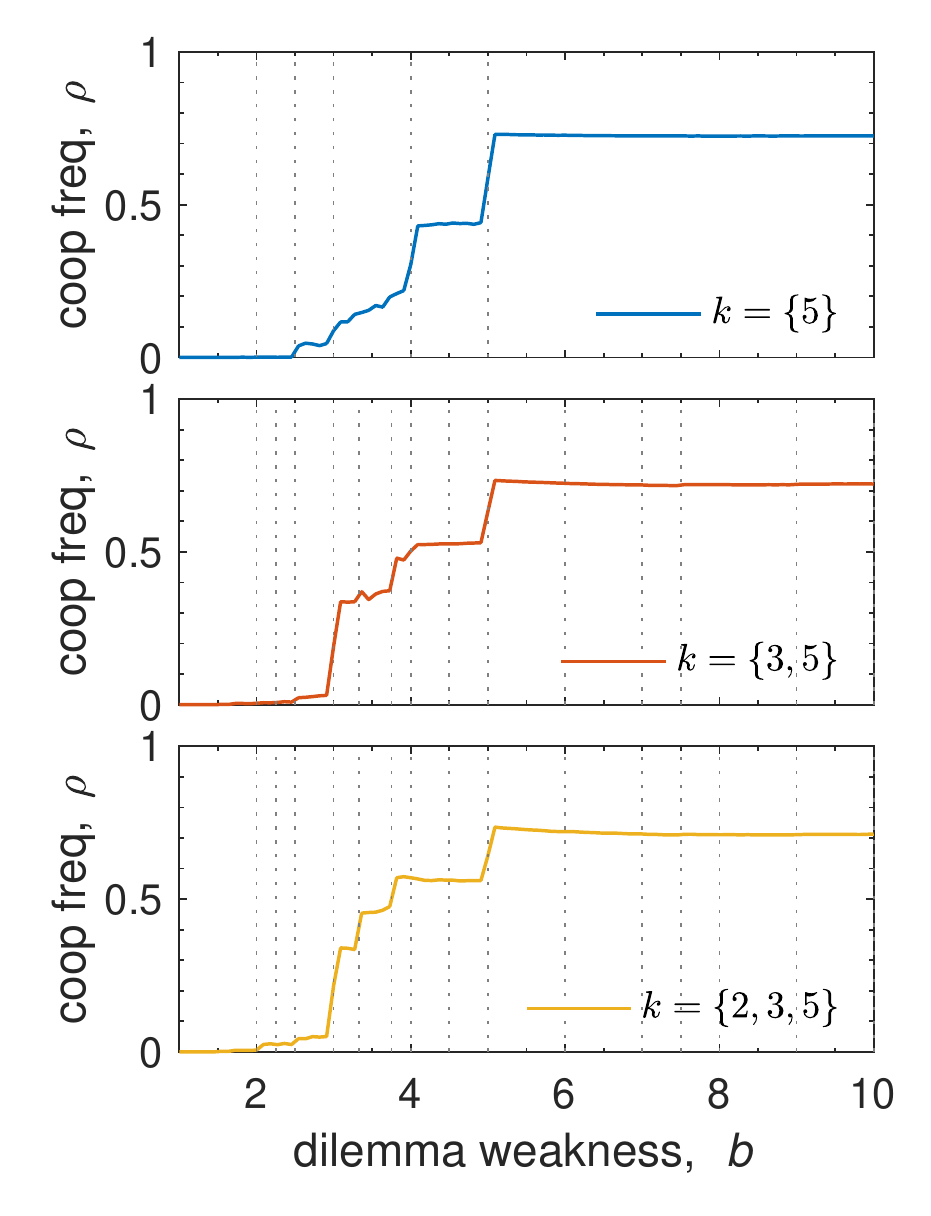}
    \caption{{ Dependence of the conjecture on the set of different degrees present in the network. } Average cooperation frequency $\rho$ as a function of the dilemma weakness $b$ for random regular graphs of size $N=600$ and different degree distributions: (a)  $k=\{5\}$, (b)  $k=\{3,5\}$ ($N_3=N_5=300$), (c) $k=\{2,3,5\}$ ($N_2=N_3=N_5=200$). The vertical grey dotted lines correspond to the $b^*$ values computed with Eq. (\ref{linkconjec}). The curves are averaged over 100 network realizations. Other parameter values: $\alpha=0.05$, $T=10^4$.}
    \label{fig1}
\end{figure}
We test our conjecture by running an extensive set of simulations of PD competition in a wide range of structural and dynamical conditions. Unless otherwise indicated, all the simulations were carried out for a population of $N=600$ players using {\it link strategy}, with initially equal densities of cooperation and defection randomly distributed along with the links. Each simulation evolves over 10,000 time steps, and cooperation frequencies are averaged in the last 2,500 time steps. At each time step, all players update their strategies synchronously \cite{nowak_n92b,huberman_pnas93}. Average cooperation frequency is monitored as a function of $b$, and each point is an ensemble average over 100 network realisations.

\begin{figure}
    \centering
    \includegraphics[width=0.9\textwidth]{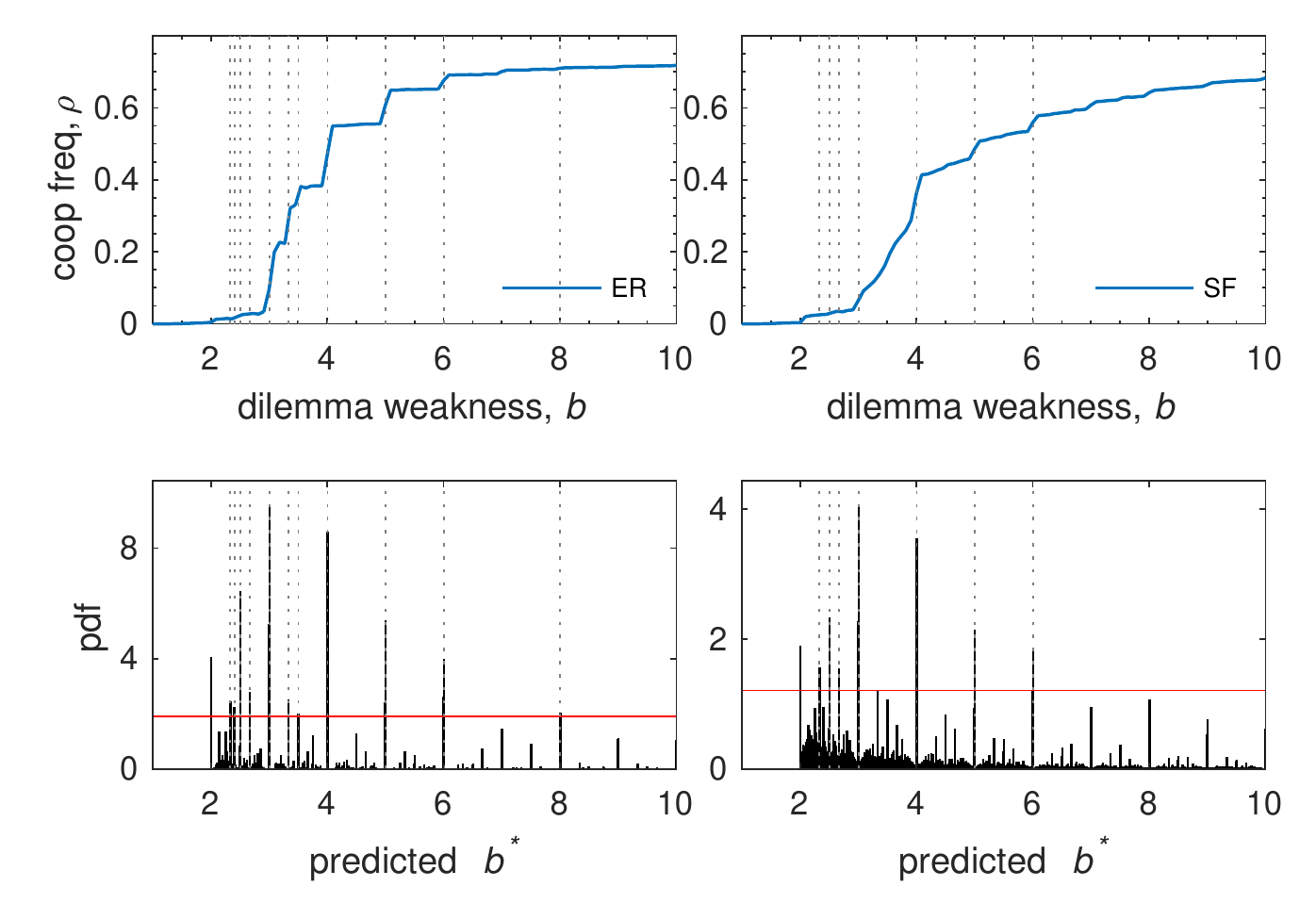}
    \caption{{ Dependence of the conjecture on the heterogeneity of the degree distribution and multiplicity of the critical game parameter. }(Top panels) Average cooperation frequency $\rho$ as a function of the dilemma weakness $b$ for ER (left) and SF (right) graphs of size $N=600$ and $\langle k\rangle=4$. Each curve is the average of 100 simulations of the same type of network. (Bottom panels) Probability density function of the $b^*$ resulting from Eq.~(\ref{linkconjec}) for the ER (left) and SF (right) degree distributions.  Vertical dotted lines in all panels correspond to the local maxima whose prominence is above the threshold marked by the red horizontal line ($20\%$ of the maximum value).}
    \label{fig2}
\end{figure}

As the conjecture implies that the number of possible transitions depends on the set of degrees present in the network, we initiate our exploration by showing the results of simulations on random regular (RR) graphs composed of groups of equal size $N_{k}$ whose nodes have all the same connectivity $k$. These graphs are constructed using the configuration model \cite{Molloy1995}, which returns a random graph with no degree correlations consistent with a given degree distribution. Specifically, we explore in Fig.~\ref{fig1} the evolution of the cooperation frequency in RR graphs for the following degree distributions:  a unique group of $N_{k=5}=N$ nodes of degree $k={5}$ (upper panel); two equally sized groups $N_3=N_5=N/2$ of agents with degrees $k=\{3,5\}$ (middle panel); and three equally sized groups $N_2=N_3=N_5=N/3$ with degrees $k=\{2,3,5\}$ respectively (bottom panel). In addition, we have included dotted vertical lines marking the values of $b^*$ predicted by  Eq.~(\ref{linkconjec}) for the chosen degree sequences. Note that the maximum degree is the same in the three cases but, as smaller degrees are included, the cardinality of $b^*$ increases adding new locations for a potential phase transition. We observe that all significant changes in the cooperation frequency always occur at one of the possible $b^*$ values predicted by Eq.~(\ref{linkconjec}).

\begin{figure}
    \centering
    \includegraphics[width=0.9\textwidth]{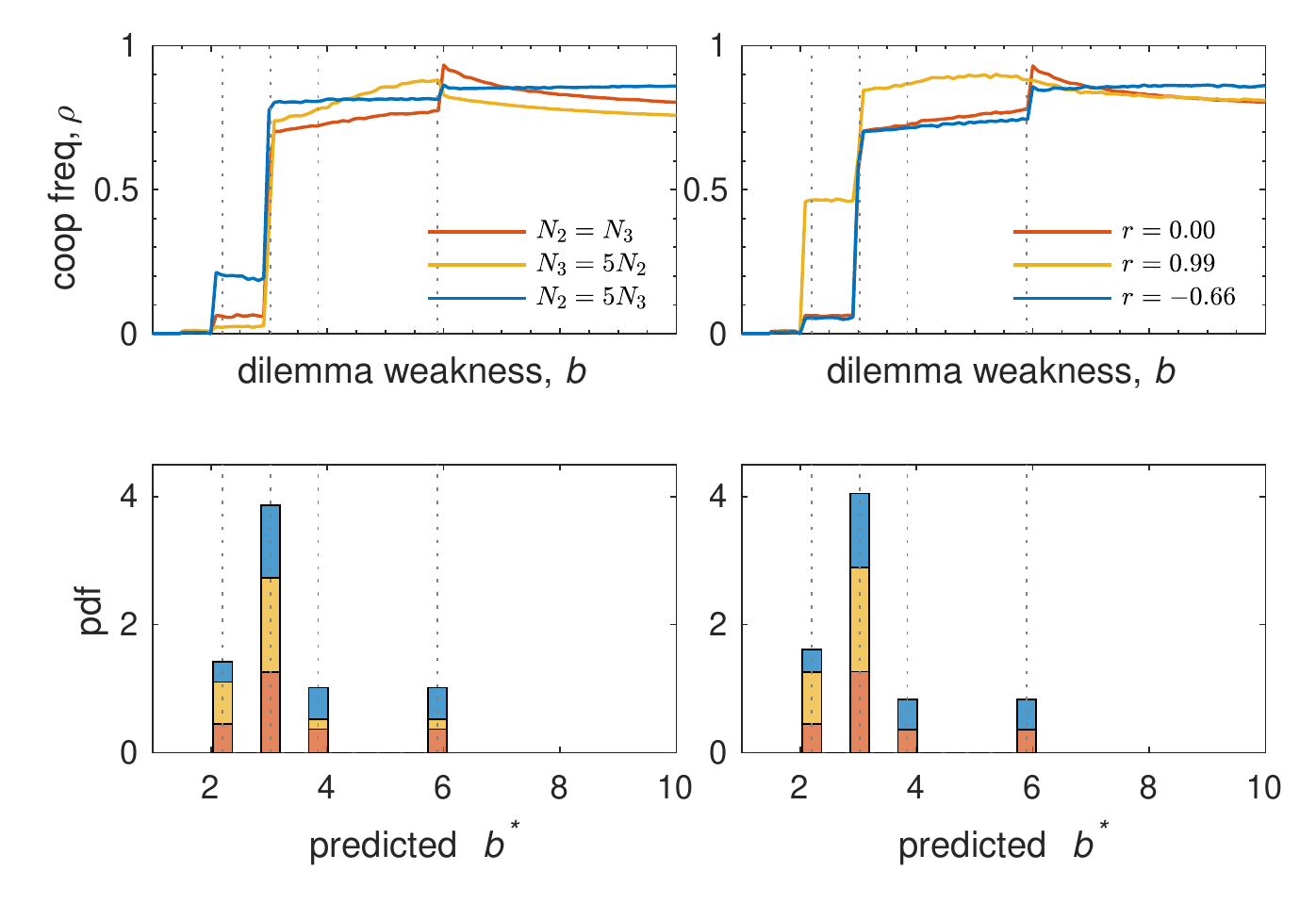}
    \caption{{ Dependence of the conjecture on the network assortativity.} (Top row) Average cooperation frequency $\rho$ versus the dilemma weakness $b$ of random regular graphs with degrees $k=\{2,3\}$ for (left) different population size and (right)  different levels of assortativity. (Bottom row) Frequency distribution of the corresponding $b^*$ values calculated considering the actual degree pairings in the networks.}
    \label{fig3}
\end{figure}

In order to inspect the role of the degree distribution  in more detail, we carried out simulations with random Erd\"os-Renyi \cite{Erdos1960} and scale-free Barab\'asi-Albert \cite{Barabasi1999}  networks, with $N=600$ and average degree $\langle k\rangle =4$. In the upper panels of Fig.~\ref{fig2} we observe that ER networks support higher values of the cooperation frequency and that in both cases, cooperation increases in a stepwise manner, with more pronounced plateaux in the ER case.
For such more complex networks, the set of possible $b^*$ values predicted by Eq.(\ref{linkconjec}) is very numerous, and one sees that not in all of them does a significant change in the cooperation frequency appears. To explain this, we consider not only the sequence of unique degrees, but the degree distribution. By knowing how often each degree $k$ appears in the network, we can calculate the multiplicity of each $b^*$ value, resulting in an excellent indicator of how probable is that a certain value $b^*$ determines a sharp variation in the cooperation frequency. In the bottom panels of Fig. \ref{fig2} we plot the histograms of the average multiplicity of the $b^*$ values in Eq.~(\ref{linkconjec}) in the degree distributions of the ER (left) and SF (right) networks.  The vertical dotted lines mark those local maxima whose prominence (height difference with respect to the local surrounding baseline) is above a given threshold (red horizontal line). Remarkably, the steps transitions observed in the cooperation frequency occur precisely at these more frequent $b^*$, as shown by the same vertical dotted lines in the top panels.

To further investigate the role played by the second order structural details we study the impact of varying the correlation degree properties of the network while keeping its degree sequence. Figure~\ref{fig3} (top left panel) reports how cooperation evolves in RR networks which include only nodes with degrees $k=\{2,3\}$, but controlling the population size $N_k$: equal size ($N_2=N_3$) and two asymmetric cases where one population is much larger than the other ($N_3=5N_2$ or $N_2=5N_3$). One clearly observes that the relevant jumps in the cooperation occur at exactly the same values of $b$, namely, $b=2,3,6$, matching the predicted $b^*$ values highlighted by the vertical dotted lines.  However, the steepness of the transitions is certainly not the same due to a difference in the  $b^*$ multiplicity caused by different degree frequencies.

A more subtle but showing case is presented in Figure~\ref{fig3} (top right panel). Here the degree distribution is preserved in all cases and $N_2=N_3=300$. Therefore, the theoretical $b^*$ multiplicity coming from the degree distribution is the same. However, in this example we control the level of degree-degree correlation measured by the associated Pearson coefficient $r$ \cite{Newman2003} (see legend in the right panel). Starting from uncorrelated networks ($r=0$), we perform the rewiring procedure of Ref.~\cite{Sendina2015} that preserves the degree distribution until the desired level of degree-degree correlation is achieved \cite{Maslov2002,Xulvi2004}. This way, one can produce graphs with very high positive correlation, in which nodes with the same degree tend to be linked among them ($r\sim 1$), and networks whose nodes have $k=2$ are more likely to be linked to nodes with $k=3$ ($r\sim -0.7$). The clear differences in the transitions (in particular for positively correlated networks) indicates that, since the calculation of $b^*$ involves all possible pairs of degrees, in correlated networks the real multiplicity could be shifted from the neutral one determined only by the degree distribution. Indeed, the bottom panels in Fig.~\ref{fig3} show the histograms for the actual multiplicity of $b^*> k_{\rm min}$ considering the real pairings in the adjacency matrices, revealing that the height of the bars differs in each case, and they provide a closer predictor of a phase transition in the game.

\begin{figure}
    \centering
    \includegraphics[width=0.9\textwidth]{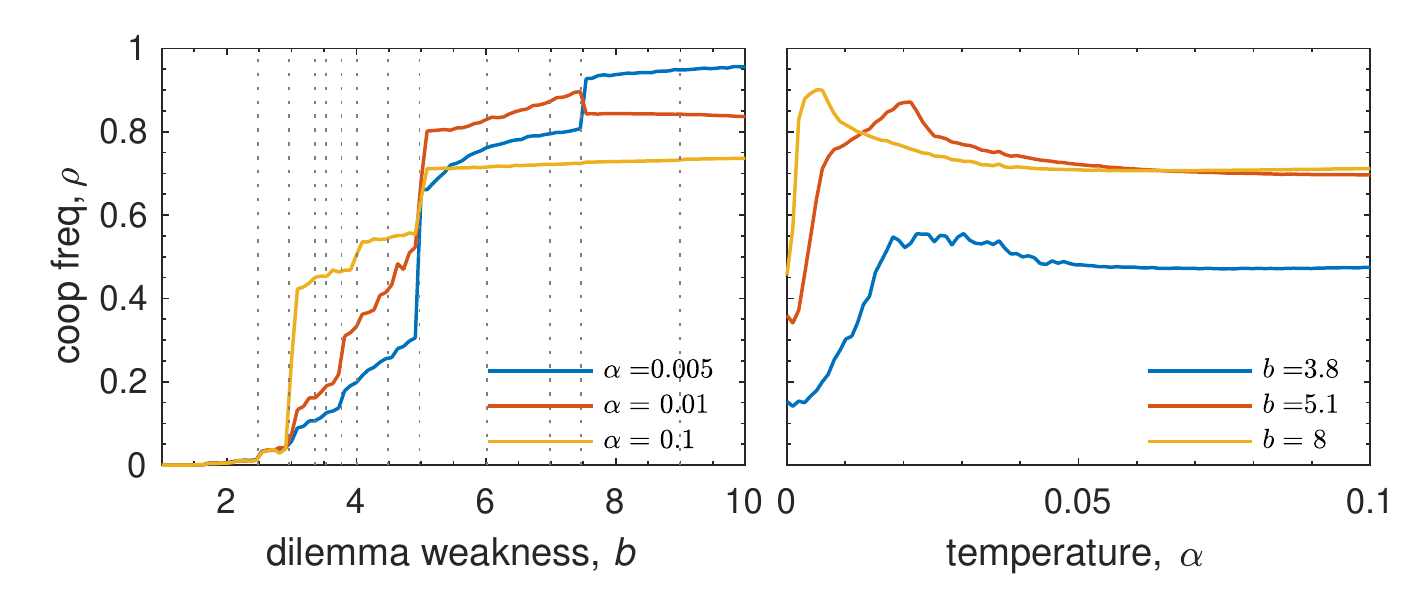}
    \caption{{ Assessment of the impact of the temperature on the conjecture's predictive performance. } Cooperation dependence  on the temperature $\alpha$ for RR graphs ($k=\{3,5\}$ and $N_3=N_5$). Panel (a) shows the cooperation $\rho$ as a function of $b$ for several values of $\alpha$. Vertical dotted lines are the corresponding $b^*$ values predicted by Eq.~(\ref{linkconjec}) for the RR network. Panel (b) shows the evolution as a function of $\rho$ for values of $b$ chosen close to a transition in (a). The network size is $N=600$ and each curve is an average over 100 network realizations.}
    \label{fig5}
\end{figure}
\begin{figure}
    \centering
    \includegraphics[width=0.55\textwidth]{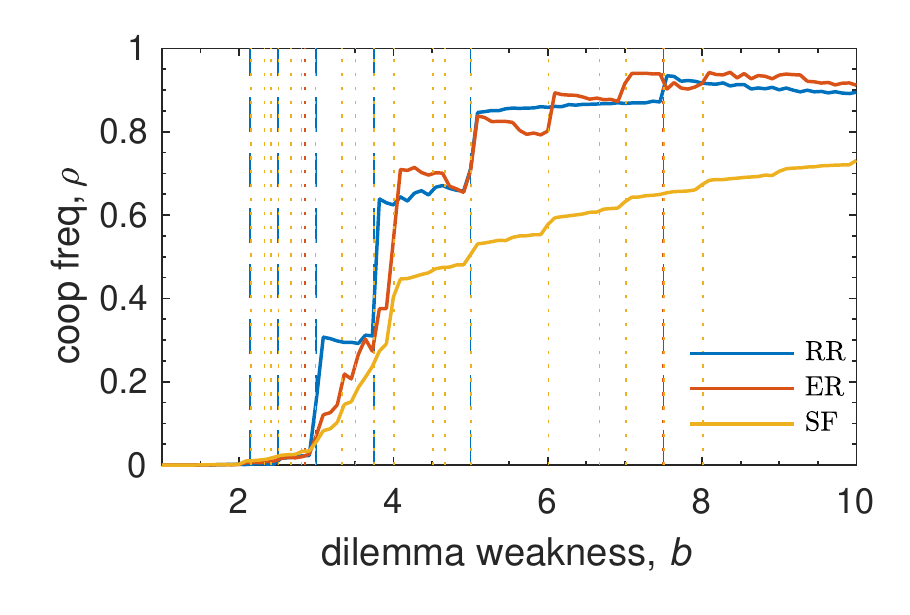}
    \caption{{ Extension of the conjecture to the prisoner's dilemma game on networks using node strategy. }Average cooperation frequency $\rho$ as a function of the dilemma weakness $b$ in a PD game using {\it node strategy} on different networks of size $N=600$ and $\langle k \rangle=4$: RR with $k=\{3,5\}$ ($N_3=N_5$), ER and SF. Curves are averaged over 100 network realizations. Vertical  lines correspond to the most frequent $b^*$ values given by Eq.~(\ref{nodeconjec}) after substituting the degree sequences of the RR (blue dashed), ER (red dotted) and SF (orange dotted) networks. Other parameter values: $\alpha=0.05$.}
    \label{fig4}
\end{figure}
Notice that our conjecture states that if there is a sharp change in the cooperation frequency, the corresponding $b$ is one of the $b^*$ values, but the reverse is not always true. This unidirectional correspondence can be verified, for instance, in Fig.~\ref{fig1}, where not all the predicted $b^*$ values (indicated by the dotted vertical lines) lead to an abrupt change in the cooperation. 
As the conjecture involves a critical phenomenon, we also have to carefully inspect the role of the stochastic part of the dynamics, here represented by the temperature parameter $\alpha$ which controls the shape of the probability function for the strategy update. To this end, in Fig.~\ref{fig5} we show the evolution of cooperation in RR graphs with $k=\{3,5\}$ ($N_3=N_5$ as in Fig.~\ref{fig1}(b)) for several values of the temperature. In addition, we also include the corresponding $b^*$ values as vertical dotted lines. One easily sees substantial changes in cooperation always coinciding with one of the vertical lines but, depending on the temperature, different $b^*$ are selected, and also differences in the height of the transition are observed. For instance, while lower temperatures support cooperation changes around $b^*=7.5$,  for $\alpha=0.1$, cooperation does not evolve. Temperature dependence of the cooperation is shown on the right panel for values of $b$ close to the major transitions observed in the left panel. The results suggest the existence of an optimal temperature that maximises cooperation, with a peak that shifts to lower temperatures as $b$ increases.

The overall scenario just described in the case of link strategy remains qualitatively unchanged for the simpler case of node strategy, considered as a particular case in Eq.~(\ref{nodeconjec}). Figure \ref{fig4}  shows the cooperation curves for node strategy in RR graphs with $k=\{3,5\}$ and ER and SF graphs with the same average degree $\langle k\rangle =4$. In general, as compared with Fig.~\ref{fig1}(b) and Figs.~\ref{fig2}(a,b), one observes that cooperation frequencies reach higher values with node strategy games. Although Eq.~(\ref{nodeconjec}) predicts a smaller set of  $b^*$ than Eq.(\ref{linkconjec}) for the same degree sequence, it is still able to capture the major transitions as shown by the vertical lines corresponding to the $b^*$ with higher prominence in their frequency distribution. For example, if we closely compare the transition to cooperation in the RR graphs using link and node strategies, Fig.~\ref{fig1}(b) and Fig.\ref{fig4} (blue line), major changes take place at $b=3.0,3.75$ and $5$, as predicted by both schemes.

For the sake of comparison with other classic studies \cite{santos_prl05,szolnoki_pa08}, we also numerically check the conjecture obtained in Eq.~(\ref{eq:bstarPnc}) for the unpunished case $P'$ introduced at the end of Sec. \ref{framework}. Furthermore, in this case the system evolution is carried out by implementing Monte Carlo simulations such that whenever the payoffs of two  engaged $u$ and $v$ { randomly selected} players verify $g_v>g_u$, then player $u$ imitates $v$'s strategy with a probability $p=\frac{D}{2\Gamma}$ for $D<2\Gamma$ and $p=1$ otherwise, being $D=(g_v-g_u)/(b\max(k_u,k_v))$ a normalized payoff difference. Notice that the slope $\Gamma$ is playing a similar functional role than $\alpha$ in the sigmoidal function of Eq.~(\ref{prob}) and that  $\Gamma=0.5$ corresponds to the case reported in Refs.~\cite{Santos2005,szolnoki_pa08}.

\begin{figure}
    \centering
    \includegraphics[width=0.9\textwidth]{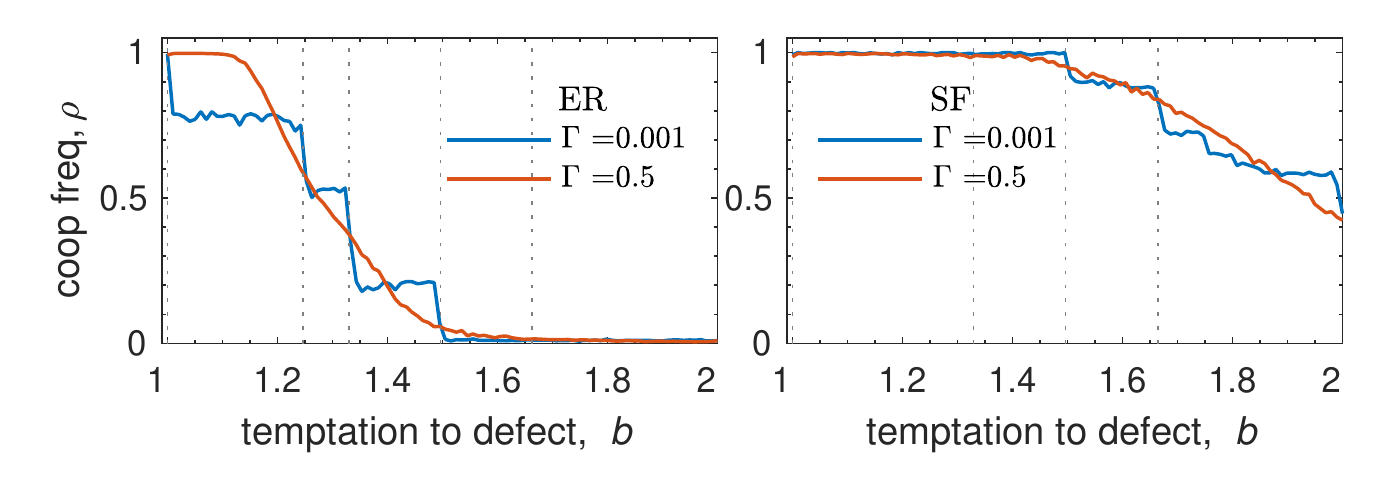}
    \caption{{ Extension of the conjecture to the prisoner's dilemma game with no cost and non synchronous strategy update of randomly selected agents.} Average cooperation frequency $\rho$ as a function of the temptation to defect (or dilemma strength) $b$ for ER (left) and SF (right) graphs of size $N=500$ and $\langle k\rangle=4$. Each curve is the average of 100 simulations of the same type of network implementing a Monte Carlo method following a transition probability given by $p=\frac{D}{2\Gamma}$ for $0\le D\le 2\Gamma$ and $p=1$ for $D>2\Gamma$. Equilibrium frequencies are obtained by averaging in the last 2,500 generations before the transient time of 10,000 generations. 
    Vertical dotted lines in both panels correspond to those $b^*$ values from Eq.~(\ref{eq:bstarPnc}) whose pdf is above 10\%  and 5\%   for ER and SF networks respectively. }
    \label{fig6}
\end{figure}
 Figure \ref{fig6} shows how the cooperation evolves as a function of the game parameter $b$  for the dilemma with no cost for both ER and SF networks and two values of the parameter $\Gamma$ quantifying the uncertainty of strategy adoptions. We observe that cooperation evolves in a continuous way for $\Gamma=0.5$ (transition probability is directly the normalised payoff difference as used in Ref. \cite{santos_prl05}) and, phase transitions from pure cooperation to a mixed state (and from a mixed state to complete defection for ER topologies) are perfectly predicted by those $b^*$ values (marked as dotted vertical lines) in Eq.~(\ref{eq:bstarPnc}) whose probability density function is above a given threshold. One sees that reducing the value of $\Gamma$ (i.e. increasing the likelihood of a player to change its strategy even for very low payoff differences) breaks the continuous evolution of the cooperation into a series of plateaux at different mixed levels of defection and cooperation.

\section{Discussion}

Cooperation is much more widespread in nature than the Darwinian premise of `only the fittest survive' might suggest. Why and under which conditions cooperation thrives is, therefore, an evergreen subject across the social and natural sciences \cite{axelrod_84,henrich2007humans,nowak_11,rand_tcs13,kraft_cobs15,perc_pr17}. Evolutionary game theory is traditionally used to formalise the problem mathematically with social dilemmas~\cite{weibull_95,hofbauer_98,nowak_06}, and networks are commonly used as the backbone for the simulation of these games. We have here shown that the structure of the former in terms of the unique sequence of degrees predict the outcome in the latter, particularly in terms of the game parameters at which significant shifts in the level of cooperation can be expected. In particular, we have proposed a conjecture for cooperation transitions in arbitrary networks, including phase transitions from absorbing to mixed strategy phases. 

Results based on the evolutionary prisoner's dilemma game on random and scale-free networks of various sizes demonstrate the effectiveness of the conjecture. { Indeed,   since the conjecture is based on the different degrees present in the network, the network size does not modify the set of game parameters critical to produce shifts in the cooperation frequency.} We have reported good agreements between the predictions and the results obtained with either a synchronous and asynchronous update of strategies, thus providing a viable alternative to computationally expensive and often time-consuming simulations. We have shown that beyond the main qualitative information provided by the unique sequence of degrees, the degree distribution and the possible second order structural correlations provide further quantitative details about the transitions.  
{ Although we have explored the predictive power of the conjecture only for the prisioner's dilemma, it can be generalized to any social dilemma as the conjecture just operates on the payoff matrix and the critical game parameters are extracted only from the condition that the payoff difference between two agents is zero.}
Despite the success, we also note the requirements for the conjecture to work best. Namely, the presence of high stochasticity levels in the probability of strategy update may compromise the emergence of sharp transitions, which can explain the fact that these have not been previously observed.

In this light, the conjecture holds the promise of the optimisation of large-scale simulations of evolutionary processes on complex networks, not just related to cooperation as a particular example of prosocial behaviour, but more generally for the broader class of moral behaviour where similar models are often employed~\cite{capraro_jrsi21}. By providing a fast and straightforward way to estimate the outcome of evolutionary processes on arbitrary networks and a wide class of games without the need for actual simulations, the conjecture should prove helpful in a wide range of research fields, further strengthening the role of physics and network science within them.

I.S.N. and I.L.  acknowledge  support from the Ministerio de Econom\'ia, Industria y Competitividad of Spain under project FIS2017-84151-P and the Ministerio de Ciencia e Innovaci\'on under project PID2020-113737GB-I00. They also acknowledge the computational resources and assistance provided by CRESCO, the supercomputing center of ENEA in Portici, Italy. M.P. was supported by the Slovenian Research Agency (Grant Nos. P1-0403 and J1-2457).

\bigskip
\bigskip
\section*{References}
\bibliographystyle{iopart-num}
\bibliography{references}
\end{document}